\documentclass[a4paper,twoside]{article}

\usepackage{epsfig}
\usepackage{subfigure}
\usepackage{calc}
\usepackage{amssymb}
\usepackage{amstext}
\usepackage{amsmath}
\usepackage{amsthm}
\usepackage{multicol}
\usepackage{pslatex}
\usepackage{apalike}
\usepackage{enumitem}
\usepackage{url}
\usepackage{amsmath}
\usepackage{csquotes}
\usepackage{SCITEPRESS}     
\usepackage[T1]{fontenc}

\begin{document}

\title{DTATG: An Automatic Title Generator Based on Dependency Trees}

\author{\authorname{Liqun Shao and Jie Wang}
\affiliation{Department of Computer Science, University of Massachusetts, One University Avenue, 01854, Lowell, MA, U.S.A.}
\email{shaoliqun89@gmail.com, wang@cs.uml.edu}
}


\abstract{We study automatic title generation for a given block of text
and present a method called DTATG to generate titles. DTATG
first extracts a small number of central sentences that convey the main meanings of the text
and are in a suitable structure for conversion into a title. 
DTATG then constructs a dependency tree for each of these sentences and removes
certain branches using a Dependency Tree Compression Model we devise.
We also devise a title test to determine if a sentence can be used as a title.
If a trimmed sentence passes the title test, then it becomes a title candidate.
DTATG selects the title candidate with the highest ranking score as the final title.
Our experiments showed that DTATG can generate adequate titles. 
We also showed that DTATG-generated titles have higher F1 scores than 
those generated by the previous methods.}

\onecolumn \maketitle \normalsize \vfill

\section{\uppercase{Introduction}}
\label{sec:introduction}

\noindent An adequate title for a given block of text must convey succinctly the central meanings of the text.
A good title must also be catchy. 
Writers would typically go through multiple rounds of revisions to come up with a satisfactory title. 
Automatic title generation (ATG) for a given block of text aims to generate a 
title comparable to a title composed by humans.
For simplicity, we will simply call a block of text a document. 

We approach ATG in two phases. In the first phase we search for a central sentence that is ``close'' to being a title in the following sense: (1) It captures the central meanings of the text; (2) It is in a form that can be converted into a title without too much effort.
We treat such a sentence as a
title candidate. In the second phase we compress this title candidate into the final title.
More specifically, during the first phase, we extract a few central sentences that
capture the main meanings of the document.
We do so using keyword-extraction algorithms to extract keywords and rank these sentences based on the
number of keywords it contains. During the second phase,
we construct a set of rules to select a central sentence in a suitable form. We then
construct a dependency tree for each central sentence using a dependency parser. We
trim possible branches according to a set of empirical rules we devise, and
 to form the final title.
We name our system Dependency-Tree Automatic Title Generator (DTATG).

If a document that already has a title, and we are asked to
generate an alternative title for the document, then we can augment the above algorithm by generating title candidates in the first phase as follows: After
computing central sentences, we further compute
the similarities of these sentences to the exiting title, and select the ones with higher similarities.

ATG may also be used in applications where we are asked to compose an article on a given title. We may use ATG to generate a title of what we have written and compute the similarity of the generated title to the given title. The article would be considered on the right track if the two titles are sufficiently similar.
In applications where we may need to paraphrase the original document, we can also use ATG to generate a new title, which may be more suitable to the new document.

We may also use ATG to generate a label for a topic in a document, where a topic is represented by a set of words relevant to that topic.
For example, given a document in a corpus, we may first use the LDA algorithm \cite{Blei} to identify the topics contained in the document, where each topic is a set of words, and then
use ATG to generate a short phrase or a short sentence to label each set.

Keyword extractions and sentence compressions are basic ATG building blocks.
Keywords provide a compact representation of the content of a document, where a keyword is a sequence of one or more words. Hence, a sentence having a larger number of keywords is expected to better convey the main meanings of a document. A popular method to identify keywords is to use the TF-IDF measure. Given a document in a corpus, a term in the document with a higher TF-IDF value implies that it appears more frequently in the document, and less frequently in the remaining documents. However,
for a corpus of a small number of documents, the TF-IDF value of a keyword would almost equal to its frequency.

The {\em Word Co-occurrence\/} (WCO) method~\cite{Matsuo:04} is a better method, which
applies to a single document without a corpus.
WCO first extracts frequent terms from the document, and
then collects a set of of word pairs co-occurring in the same sentences (sentences include titles and subtitles), where one of the words is a frequent term.
If term $t$ co-occurs frequently with a subset of frequent terms, then it is likely to be a keyword.
The authors of WCO showed that WCO offers comparable performance to TF-IDF without the presence of a corpus.

The {\em Rapid Automatic Keyword Extraction\/} (RAKE) algorithm~\cite{Rose:10} is another keyword extraction method
based on word pair co-occurrences. In particular, RAKE first divides the document
into words and phrases using predetermined word delimiters, phrase delimiters, and positions of stop words.
RAKE then computes a weighted graph, where each word is a node, and a pair of words are connected with weight $n$
if they co-occur $n$ times. RAKE then assigns a score to each keyword candidate, which is the summation of scores of words contained in the keyword.
Word scores may be calculated by word frequency, word degree, or ratio of degree to frequency. Keyword candidates with top scores are then
selected as keywords for the document. RAKE is superior over WCO in the following aspects: It is simpler and achieves a higher precision rate and about the same recall rate.
We will use RAKE to extract keywords.

Early title generation methods include Naive Bayesian with limited vocabulary, Naive Bayesian with full vocabulary, Term frequency and inverse document frequency (TF-IDF), K-nearest neighbor, and Iterative Expectation Maximization. These methods, however, only generate an unordered set of keywords as a title without
concerning syntax.
Using the F1 score metric as a base for comparisons, it was shown through extensive experiments
that the TF-IDF title generation method has the best performance over the other five methods \cite{Jin and Hauptmann:01}.

For practical purposes we would like to generate syntactically correct titles.
Recent methods used
sentence trimming to convert a title candidate into a shorter sentence or phrase, while trying to maintain syntactic correctness.
Sentence trimming has been studied in recent years and has met with certain success.
For example, Knight and Marcu~\cite{Knight and Marcu:02} and Turner and Charniak~\cite{Turner and Charniak:05} used a language model (e.g., the trigram model) to trim sentences.
Vandegehinste and Pan~\cite{Vandegehinste and Pan:04} used context-free grammar (CFG) trees to trim a sentence to generate subtitle candidates with appropriate pronunciations for the deaf and hard-of-hearing people.
They used a spoken Dutch corpus for evaluation. More recently,
Zhang et al.~\cite{Zhang:13} used sentence trimming on the Chinese text to generate titles, also through CFG trees.
We note that CFG is ambiguous and the complexity of constructing CFG trees is high.

Dependency trees are recent development in natural language processing with a number of advantages. For example, dependency trees offer better syntactic representations of sentences \cite{Sylvain:12} and they are easier to work with. These advantages motivated us to explore automatic title generation using dependency trees. We present the first such algorithm.


Our approach has the following major differences from the previous approaches:
\begin{enumerate}[nolistsep]
\item We use RAKE to generate keywords and define a better measure to select central sentences.
\item We use dependency grammar to construct a dependency tree for each title candidate for trimming.
\item We construct a set of empirical rules to generate titles.
\end{enumerate}
We show that, through experiments, DTATG generates titles comparable to titles generated by human writers.
In addition to this evaluation, we also evaluate the F1 scores and show that, through experiments, DTATG
is superior over the TF-IDF method (see Section \ref{comparison}).

The remainder of this paper is organized as follows: In Section \ref{sec:title} we first provide an overview of DTATG. We then explain DTATG in detail, including extraction of central sentences, dependency parsing, and the dependency tree compression model. In Section \ref{sec:setup} we describe our experiment evaluation setups and present results from our experiments. We conclude the paper in Section \ref{sec:conclusion}.

\section{\uppercase{Detail Description of DTATG}}
\label{sec:title}
\begin{figure*}[t]
\includegraphics[width=6in]{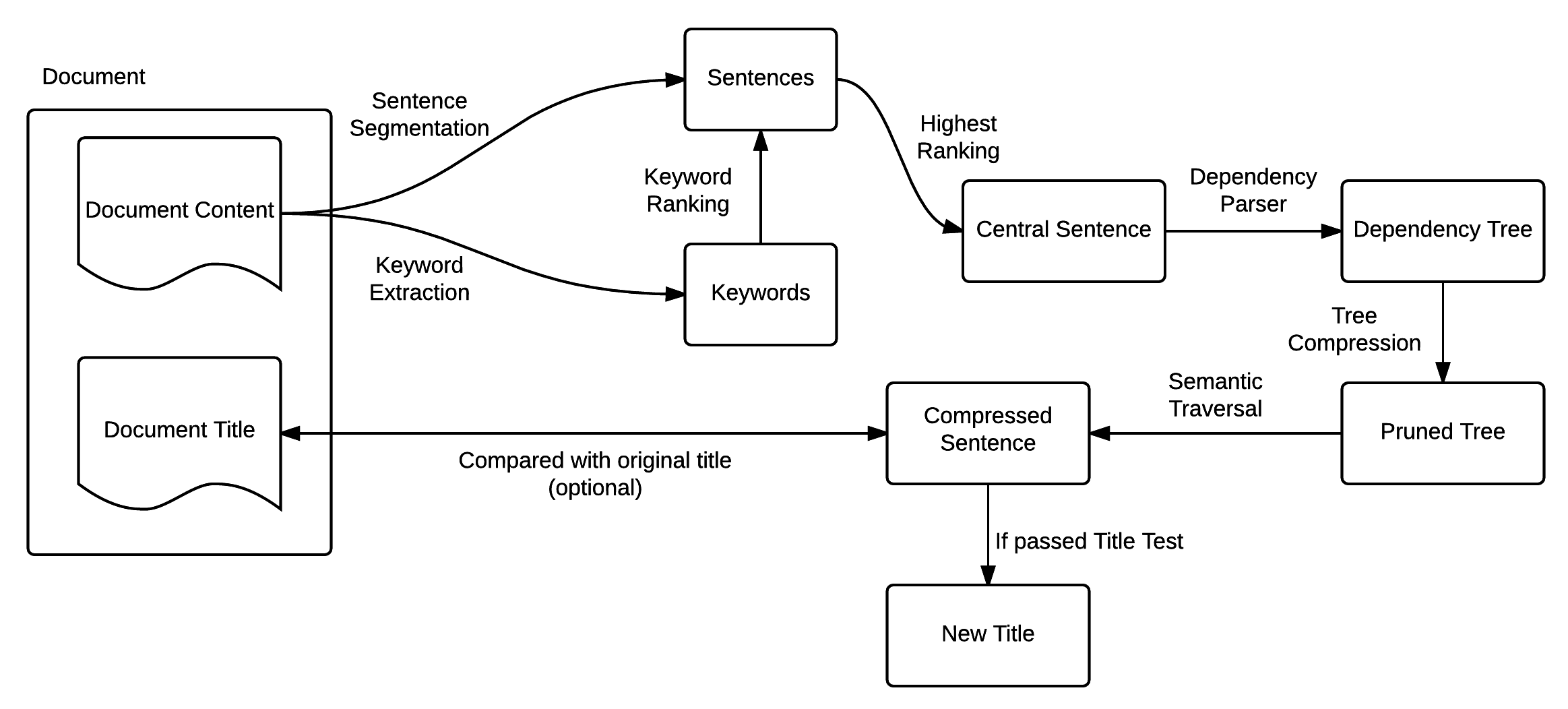}
\caption{DTATG system framework.}
\label{fig:fig1}
\end{figure*}

\noindent The DTATG system framework to generate a title for a given document is shown in Figure \ref{fig:fig1}, where we assume that each document already has a title
for comparison.
If a document does not have a title, then this comparison will not be executed.
Given a document, DTATG generates a title as follows:
\begin{enumerate}[nolistsep]
\item Extract keywords.
\item Use sentence segmentation to obtain sentences and rank each sentence using a suitable measure based on the number of keywords it contains. Select a fixed number of
sentences with the highest rankings as the central sentences. In general, selecting three central sentences would be sufficient.
\item Construct a dependency tree for each central sentence using a dependency parser, starting from the sentence with the highest ranking. In particular, we use Stanford University's open-source tool\textemdash   Stanford Dependency Parser\textemdash as our dependency parser.
\item Remove certain branches of the dependency tree based on a set of empirical rules we devise to compress the sentence.
\item If the trimmed sentence passes the title test we devise, then output it as the title.
%
\end{enumerate}

However, we note that not all documents have central sentences. For example, in a document that describes a number of items to be avoided, it
may simply state that ``The following items should be avoided:'' followed by a list of items. In this document, any sentence that describes
an item does not include the keyword ``avoided'', and the sentence that contains this keyword does not contain any keywords for describing any item.
Thus, none of the sentence in the document can represent the central meaning of the document, that is, none of the sentences can specify
what items should be avoided.

For convenience, we call documents with central sentences \textbf{type-1 documents} and documents without central sentences \textbf{type-2 documents}.
While DTATG may apply to both types of documents, it performs better for type-1 documents.


\subsection{Central sentence extraction}
\label{central}
\noindent We segment each document into sentences using a sentence-delimiter set \{., ?, !, \textbackslash n, :\}.
This set does not include comma as compared to sentence delimiters used by other researchers, for
we want to obtain a complete sentence.
From experience we should select only consider sentences with at most 25 words for title candidates.

We first use RAKE~\cite{Rose:10} to identify keywords, where RAKE assigns each keyword with a positive numerical score.
Next we will compute the ranking of each sentence to reflect the importance of the sentence based on the keywords it contains.
Assume that a sentence contains $n$ keywords $w_1, \cdots, w_n$, and $w_i$ has a positive score $s_i$.
An obvious method to define the rank of the sentence is to simply sum up the scores of all the keywords contained in it as follows:
\begin{equation} \label{rank2}
\mbox{Rank}_1 = \sum_{i=1}^n s_i.
\end{equation}
However, we observe that 
using $\mbox{Rank}_1$
we may end up giving a sentence containing a larger number of keywords of lower scores a higher rank than a sentence
containing fewer keywords of higher scores.
For example, suppose that sentence $S_1$ contains two keywords with scores of $(4,5)$
and sentence $S_2$ contains four keywords with scores of $(2, 2, 3, 3)$. Then
$S_2$ has a higher $\mbox{Rank}_1$ ranking than $S_1$. This contradicts to the common sense that
$S_1$ would be more relevant to the central meanings.

Thus, we would want keywords of higher scores to carry more weights, so that
the ranking of a sentence
with smaller number of keywords of much higher scores is higher than the ranking of a sentence with larger number of keywords of much lower scores.
To achieve this, we use the power of 2 to amplify the scores, giving rise to the following measure:

%
\begin{equation} \label{rank1}
\mbox{Rank}_2 = \sum_{i=1}^{n}2^{s_{i}}.
\end{equation}

In the example above, it is easy to see that 
$S_1$ has a higher $\mbox{Rank}_2$ ranking. On the other hand, if $S_2$ contains three keywords each with a score of 4, then
we would consider $S_1$ and $S_2$ equally important. The $\mbox{Rank}_2$ measure ensures this effect.
Moreover, if $S_2$ contains four keywords with scores of $(1,4,4,4)$, then we would want $S_2$ to have a higher ranking than $S_1$. Again, the $\mbox{Rank}_2$ measure also ensures this effect. 

Empirical results indicate that $\mbox{Rank}_2$ is a better measure for ranking sentences. For example, 
Table \ref{central} shows two central sentences obtained using $\mbox{Rank}_1$ and $\mbox{Rank}_2$ from a news article as an example, both having the highest score under the respective measure.
It is evident that, compared with the original title, the central sentence obtained using $\mbox{Rank}_2$ presents a better choice.

\begin{table}[h]
\begin{center}
\caption{\label{central} Examples of central sentence extraction}
\begin{tabular}{r l}
\bf Original title & Bad e-mail habits sustains spam \\
\bf $\mathbf{Rank_1}$ & People must resist their basic \\
& instincts to buy from spam mails \\
\bf $\mathbf{Rank_2}$ & One in ten users have bought \\
& products advertised in junk mail
\end{tabular}
\end{center}
\end{table}

DTATG uses $\mbox{Rank}_2$ to select central sentences as follows:
Order sentences first by rank and then by the sentence length. In other words, we first select the sentence with the highest rank and if two sentences have
the highest rank, we select the shorter one. 
Our experiments indicate that this method of selecting central sentences improves performance.

The sentence we selected may have multiple clauses separated by commas. If the sentence has more than two clauses, we keep the two clauses that have the highest ranking and separate them using a space. This makes it easier to perform sentence compression in the next step.

To generate suitable sentences for trimming, we apply the following empirical rules to
avoid inappropriate sentences. 
\begin{itemize}[nolistsep]
\item Discard sentences that contain a pronoun subject.
\item Discard sentences that contain predicate verbs in \{am, is, are\}.
\end{itemize}

\subsection{Dependency trees}
\label{ssec:rules}

\begin{figure}[h]
\includegraphics[width=3in]{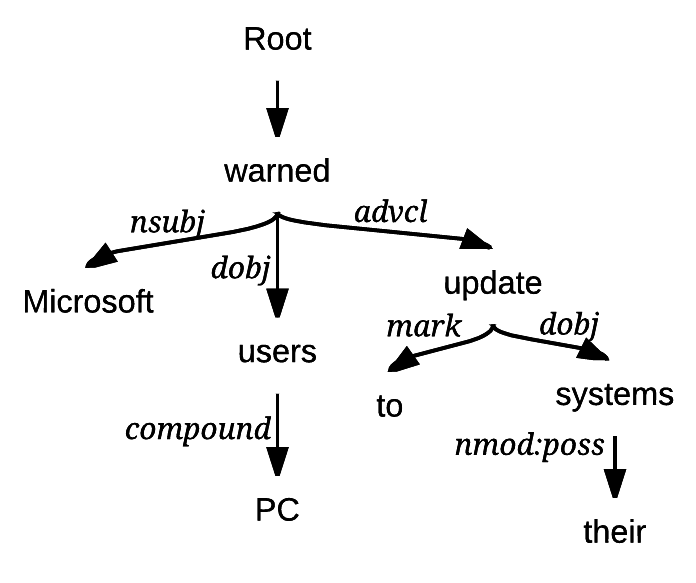}
\caption{The dependency tree of sentence ``Microsoft warned PC users to update their systems''}
\label{fig:fig2}
\end{figure}

We use the Stanford Dependency Parser (SDP), available at \url{http://nlp.stanford.edu/software/stanford-dependencies.shtml}, to obtain grammatical relations between words in a sentence. The relations are presented in triplets:
\[
\mbox{(relation, governor, dependent).}
\]
The tree is constructed by the relations between each word in the sentence. For example, {\em nsubj} stands for nominal subject, which is the syntactic subject of a clause. Universal Dependencies are documented online at \url{http://universaldependencies.org/u/dep/index.html}. For example, Figure \ref{fig:fig2} depicts a dependency tree for
the sentence ``Microsoft warned PC users to update their systems''.
In addition, we may use the Stanford Parser (SP), available at \url{http://nlp.stanford.edu/software/lex-parser.shtml}, to obtain part-of-speech tagging.

\subsection{Dependency tree compression model}
\label{ssec:compression}

We devise a dependency tree compress model (DTCM) to generate titles. First, we define
the following set of empirical rules to specify what cannot be trimmed. These rules were formed based on
our experiences from working with a large number of text documents. We refer to these rules as the \textbf{to-be-kept} rules.
%
\begin{enumerate}[nolistsep]
\item If the relation is {\em nsubj} or {\em nsubjpass}, then keep both the governor and dependent.
\item If the relation is {\em dobj} or {\em iobj}, then keep both the governor and dependent.
\item If the relation is {\em compound}, then keep both the governor and dependent.
\item If the relation is {\em root}, then keep the dependent.
\item If the relation is {\em nmod}, then keep the dependent and the preposition in \textsl{nmod}.
\item If the relation is {\em nummod}, then keep both the governor and dependent.
\end{enumerate}

We note that the remainder of the sentence after removing all the other relations not specified the to-be-kept rules can still be understandable by humans.

DTCM proceeds as follows:
\begin{enumerate}[nolistsep]
\item Traverse the entire dependency tree in preorder.
\item \label{l} For each leaf node $t$, if it is not a keyword or a part of the keyword or the relation of its parent edge is not in the set of to-be-kept rules,
    then remove node $t$.
\end{enumerate}
In general, DTCM deletes all the unnecessary branches; most of these branches would have relations such as {\em det}, {\em aux}, and {\em amod}.

\begin{figure}[h]
\includegraphics[width=3in]{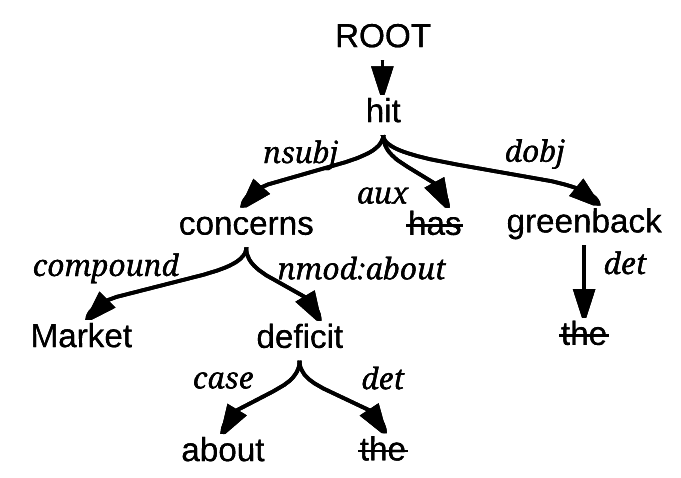}
\caption{The trimmed dependency tree using DTCM for sentence ``Market concerns about the deficit has hit the greenback.''}
\label{fig:fig3}
\end{figure}

Figure \ref{fig:fig3} shows a trimmed dependency tree using DTCM for a central sentence ``Market concerns about the deficit has hit the greenback.''
The output is ``Market concerns about deficit hit greenback'', which would be a suitable title.

\subsection{Compression rate}
\label{twice}

To transform a trimmed dependency tree into a title, we output the words in the same order of the original sentence.
Let $y$ be a compressed sentence for a sentence ${x = w_{1}w_{2}w_{3}\cdots w_{n}}$, where ${w_{i}}$ is a word.
We use \enquote{0} to mark a word to be removed and \enquote{1} to mark a word to be retained.  Then a compression rate {\em r} for a sentence is defined as follows:
\begin{equation}
r = \frac{\mbox{\# of 1s in } y}{n}.
\end{equation}
If $r$ is too large or too small, we may delete more or fewer words, respectively, to make the output more reasonable.
For example, since the length of a good title should not exceed 10 words (this is a common-sense rule), 
if a central sentence is more than 20 words and the compression rate $r$ is less than 50\% after pruning,
then we may apply DTCM again on the trimmed sentence with the hope to obtain a higher compression rate.

The following is a set of empirical rules for deletion and we refer to them as \textbf{to-be-deleted} rules. We may keep adding rules to this set.

\begin{itemize}[nolistsep]
\item Delete the first adverbial phrase and the last adverbial phrase.
\item Delete \enquote{X says} and \enquote{X said}, where X is a noun or pronoun.
\item If there are two clauses connected with \enquote{and} and the first clause has subject and verb, delete the second clause with \enquote{and}.
\item If there is a clause starting with \enquote{that} and the clause has a subject and a verb, then delete all the words before \enquote{that} (including \enquote{that}).
\item If there are more than one {\em nmod} relations next to each other, then delete all the {\em nmod} relations except the last one.
\end{itemize}

\section{TITLE TEST}
\label{sec:standards}

\noindent 
A good title must pass the \textbf{title test}, which consists of the
following three individual tests.

\noindent
\textbf{The conciseness test}.
\begin{itemize}[nolistsep]
\item A title should not exceed 15 words.
\item A title must not have clauses.
\item A title should have the following structure: \enquote{Subject + Verb + Object} or \enquote{Subject + Verb}. Subject must be specific: it can be a noun but not a pronoun.
\end{itemize}

\noindent
\textbf{The fluency test}.
A title should contain no grammatical errors.

\noindent
\textbf{The topic-relevance test}.
A title must convey at least one major meaning of the document.

DTATG uses central sentences to achieve topic relevance, and uses dependency grammar and empirical rules to achieve conciseness and fluency.
We will use the title test to evaluate the quality of the generated titles by DTATG.

\section{EVALUATIONS}
\label{sec:setup}

\noindent We evaluate DTATG on a corpus of English documents obtained from the BBC news website~\cite{Greene and Cunningham:06}. The dataset can be found at \url{http://mlg.ucd.ie/datasets/bbc.html} and \url{https://drive.google.com/open?id=0B_-6yYneQ1_8Ml91TnY0THJoYXc}, which consists of 2,225 documents in five topical areas from the year of 2004 to the year of 2005.
These five
topical areas are business, entertainment, politics, sport, and tech. The corpus provides raw text files with original articles and titles. The datasets include type-1 and type-2 documents. In our experiments, we will use type-1 articles in the business, entertainment, politics, and tech categories and type-2 articles in the sport category.

\begin{table*}[t]
\begin{center}
\caption{The average scores for DTATG-generated titles compared with the original titles} \label{results}
\begin{tabular}{c |c| c| c| c| c| c}
\hline
\bf Category & \bf Base-CON & \bf CON & \bf Base-FLU & \bf FLU & \bf Base-TR & \bf TR \\
\hline
Business & 4.84 & 4.40 & 4.76 & 4.68 & 4.72 & 4.12 \\
\hline
Entertainment & 4.80 & 4.20 & 4.68 & 4.72 & 4.76 & 4.16 \\
\hline
Politics & 4.72 & 4.16 & 4.76 & 4.52 & 4.68 & 4.04 \\
\hline
Sport & 4.80 & 4.12 & 4.60 & 4.44 & 4.40 & 4.68 \\
\hline
Tech & 4.96 & 3.84 & 4.44 & 4.32 & 4.32 & 4.12 \\
\hline
\end{tabular}
\end{center}
\end{table*}

\subsection{Experiment setup}

We carried out experiments on a computer with 2.7 GHz dual-core Intel Core i5 CPU and 8 GB memory. In our experiments, we used a Python implementation of the RAKE algorithm to extract keywords,
and the Stanford Parser to generate dependency trees, where
the source code for implementing RAKE can be found at \url{https://github.com/aneesha/RAKE} and
the Stanford Parser can be downloaded from \url{http://nlp.stanford.edu/software/lex-parser.shtml#Download}.

\subsection{Results and discussions}
\label{sec:results}

We randomly selected 50 articles in each category to evaluate. 
We asked English speakers to rank each DTATG-generated title in the category of conciseness, fluency, and topic relevance using a 5-point scale.

\subsubsection*{Rating definition on conciseness (CON)}

\begin{itemize}[nolistsep]
\item[5:] Excellent; every word in the sentence is to the point that makes a perfect title.
\item[4:] Very good; the sentence contains no redundant words that makes a very good title.
\item[3:] Good; the sentence has some redundant words, but may still be used as a title.
\item[2:] Weak; the sentence has some redundant words, but is a poor choice as a title.
\item[1:] Poor; the sentence contains a lot of redundant words and should not be used as a title at all.
\end{itemize}

\subsubsection*{Rating definition on fluency (FLU)}

\begin{itemize}[nolistsep]
\item[5:] Excellent; the sentences contains no grammatical flaws.
\item[4:] Very good; the sentences contains only minor grammatical flaws.
\item[3:] Good; the sentence contains some grammatical flaws.
\item[2:] Weak; the sentence is only partially intelligible, with serious grammatical flaws.
\item[1:] Poor; the sentence makes no sense at all.
\end{itemize}

\subsubsection*{Rating definition on topic relevance (TR)}

\begin{itemize}[nolistsep]
\item[5:] Excellent; the sentence has perfect topic relevancy.
\item[4:] Very good; the sentence has significant topic relevancy.
\item[3:] Good;  the sentence has moderate topic relevancy.
\item[2:] Weak;  the sentence has marginal topic relevancy.
\item[1:] Poor; the sentence is topic irrelevant.
\end{itemize}

\begin{table*}[t]
\begin{center}
\caption{Comparison examples of DTATG-generated titles with the original title showing the score (CON, FLU, TR)} \label{example}
\begin{tabular}{l|l}
\hline
\hline
\bf Original title  & Blogs take on the mainstream (5, 5, 4)\\
\bf Central sentence  & The blogging movement has been building up for many years. \\
\bf DTATG title & Blogging movement building up for years (5, 5, 5)\\
\hline
\bf Remark
& The DTATG title contains more information, and so we consider it better than \\
& the original title. \\
\hline
\hline
\bf Original title & Day-Lewis set for Berlin honour (5, 5, 4)\\
\bf Central sentence & Japan's oldest film studio will also be honoured along with Day-Lewis. \\
\bf DTATG title & Japan's oldest film studio honoured with Day-Lewis (5, 5, 5)\\
\hline
\bf Remark
& The original article talked about both Day-Lewis's and Japan's oldest film studio's \\
& contributions. The original title only mentioned Day-Lewis. The DTATG title \\
& mentioned both, and so we would consider it a better title. \\
\hline
\hline
\bf Original title & Scots smoking ban details set out (5, 5, 5)\\
\bf Central sentence & A comprehensive ban on smoking in all enclosed public places in Scotland. \\
\bf DTATG title & Comprehensive ban on smoking in public places in Scotland (5, 5, 5)\\
\hline
\bf Remark
& The DTATG title is as good as the original title made by a professional writer. \\
\hline
\hline
\bf Original title  & Xbox 2 may be unveiled in summer (5, 5, 5)  \\
\bf Central sentence  & Since its launch, Microsoft has sold 19.9 million units worldwide. \\
\bf DTATG title   & Microsoft sold 19.9 million units worldwide (5, 5, 4)\\
\hline
\bf Remark
& The DTATG title is missing Xbox as the name of the units sold, and so it is not \\
& a very good title. \\
\hline
\hline
\end{tabular}
\end{center}

\end{table*}

\subsubsection*{Evaluation results}

\noindent
Each evaluation category was evaluated by 5 human evaluators. We then average their scores. On the conciseness and fluency evaluation, we asked the evaluators to compare the original titles with DTATG-generated titles. On the topic relevance evaluation, we asked the evaluators to make a subjective judgment on how much the DTATG-generated titles represent the main topics of the article. For this evaluation, the evaluators needed to read the original articles.
We note that most articles in the sport category are type-2 documents. In these documents we simply used the first sentence as the central sentence.

\begin{figure}[h]
\includegraphics[width=3in]{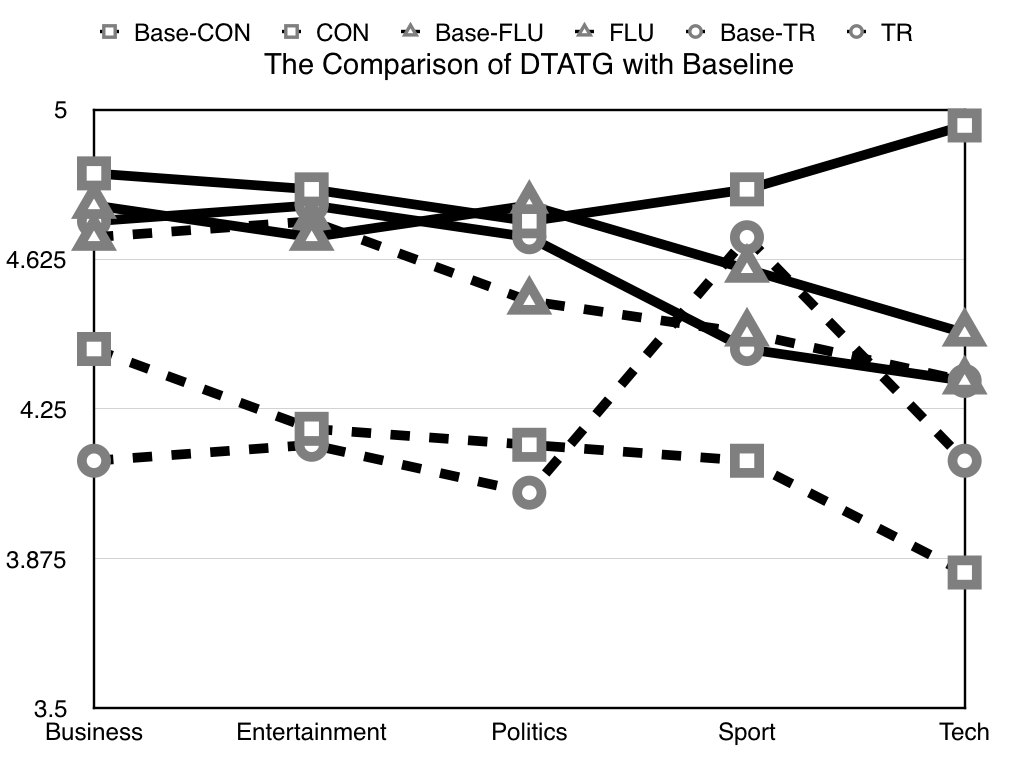}
\caption{Line graph of DTATG-generated titles and original titles on CON, FLU, and TR}
\label{fig:fig4}
\end{figure}

To better assess the quality of the DTATG results, we also ask evaluators to evaluate the original titles using the same rules, and we use
the original as the comparison baseline.
Results are given in Table \ref{results} and Figure \ref{fig:fig4} for a better visual. In Figure \ref{fig:fig4}, the dotted lines represent the scores of titles generated by DTATG and the solid lines represent the scores of original titles.

From Table \ref{results} we can see that the titles generated by DTATG have scores over 4 on fluency and topic relevance on all categories.
Moreover, most of the titles are concise except in the tech category. The reason why the generated titles for the tech articles are not as concise is because these articles
often contain long technology names or long company names, which make titles longer. Nevertheless, DTATG in general does reasonably well on all evaluation categories of conciseness, fluency, and topic relevance. DTATG-generated titles are understandable and acceptable by human evaluators. From Figure \ref{fig:fig4}, we can easily tell that titles generated by DTATG are close to baselines. Titles of FLU in Entertainment and TR in Sport categories are better than the baseline, which means they are even better than the original titles because they provide more information (see Table \ref{example}).

Detailed information of these examples,
including the original title, the DTATG-generated title, the original text, the keywords,
the central sentence, the dependency tree for the central
sentence, and the pruned branches, can be found at \url{http://119.29.118.23/title_generator/index.html}. We also implemented a system at \url{http://www.cwant.cn/title_en/} that allows users to enter text, upload a txt file, or enter an URL to generate a title automatically using DTATG.

We note that the performance of DTATG is confined by the choice of central sentences.
To produce a good title, we must have a good central sentence to begin with. For example, in the last example in Table \ref{example}, if
``Xbox'' is included in the central sentence to read ''Since its launch, Microsoft has sold 19.9 million Xbox units worldwide'',
then the DTATG-generated title would have been ``Microsoft sold 19.9 million Xbox units worldwide'', which would be a good title.
However, the original article mainly talked about the success of Xbox and only mentioned briefly that Xbox 2 would be available in May. Thus, no central sentences would be able to capture Xbox 2.

\subsection{Comparision with TF-IDF}
\label{comparison}

\begin{figure}[h]
\includegraphics[width=3in]{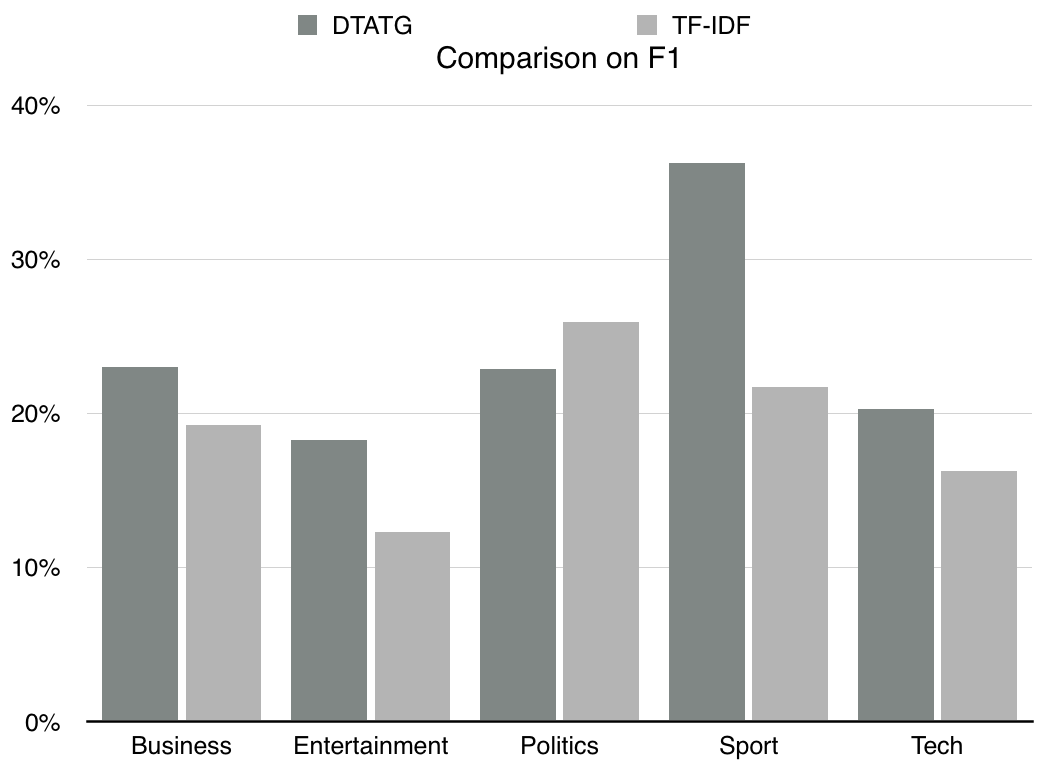}
\caption{Comparison of DTATG and TF-IDF on the F1 score}
\label{fig:fig5}
\end{figure}

For title generation using the TF-IDF method, the words in the document with highest TF-IDF would be chosen for the title.
Given a document, let T1 denote the title generated by DTATG and T2 the original title. Let
precision denote the number of identical words in T1 and T2, divided by the number of words in T1; and recall the number of identical words in T1 and T2, divided the number of words in T2. Then the F1 score is defined by
\begin{equation}
\mbox{F1} = \frac{ 2\times precision\times recall }{precision + recall}.
\end{equation}
We randomly selected 100 articles from datasets in Section \ref{sec:setup} and computed the F1 score for each article.
The average F1 scores are shown in
Figure \ref{fig:fig5}.

In four out of five categories, the titles generated by DTATG perform better than those generated by TF-IDF. In particular, for the category of sport, the average F1 score under DTATG is much higher than that under TF-IDF. In the category of politics, the average F1 score under DTATG is slightly lower than that under TF-IDF.

\section{CONCLUSION}
\label{sec:conclusion}

\noindent We presented DTATG for automatic title generation on a given block of text.
DTATG is a system combining central sentence selection and dependency tree pruning.
DTATG is unsupervised and can generate titles quickly and syntactically.
DTATG, however, is confined by the choice of central sentences.
We plan to study how to combine several central sentences
and compress them using deletion, substitution, reordering, and insertion
to generated better titles.

\section*{ACKNOWLEDGEMENTS}

\noindent This work was supported in part by a research grant from Wantology LLC. We thank Yiqi Bai, Shan (Ivory) Lu, Jing Ni, Jingwen (Jessica) Wang, Hao Zhang, and a few other people for helping us evaluate the titles generated by DTATG.


%

\vfill

\end{document}